\documentclass[12pt, journal, onecolumn, draftcls]{IEEEtran} %

\usepackage{cite}      
\usepackage{graphicx}  
\usepackage{psfrag}                            
\usepackage{subfigure} 
\usepackage{url}       
\usepackage{stfloats}  
\usepackage{amsmath}   
\usepackage{array}
\usepackage{times}
\usepackage{epsfig,graphicx,color,pstricks}
\usepackage{amsmath,amssymb,amsthm,amsbsy,amsfonts,latexsym}
\usepackage{bm}
\usepackage{psfrag}
\usepackage[mathscr]{eucal}
\usepackage{cite}
\usepackage[all]{xy}
\usepackage{subfigure}
\usepackage{graphics}

\newtheorem{thm}{Theorem}

\theoremstyle{remark}

\theoremstyle{definition}

\begin{document}

\title{Approximate Sum-Capacity of K-user Cognitive Interference Channels with Cumulative Message Sharing}
\author{
\IEEEauthorblockN{Diana Maamari, Daniela Tuninetti and Natasha Devroye,}\\
\IEEEauthorblockA{
Department of Electrical and Computer Engineering,\\
University of Illinois at Chicago, Chicago IL 60607, USA,\\
Email: {\tt dmaama2, danielat, devroye @ uic.edu}}%
}

\maketitle

\begin{abstract}
This paper considers the $K$-user cognitive interference channel with one primary and $K-1$ secondary/cognitive transmitters with 
a {\em cumulative message sharing} structure, i.e., cognitive transmitter $i\in[2:K]$ knows non-causally all messages of the users with index less than $i$. 
We propose a computable outer bound valid for any memoryless channel. We first evaluate the sum-rate outer bound for the  high-SNR linear deterministic approximation of the Gaussian noise channel. This is shown to be capacity for the 3-user channel with  arbitrary channel gains and the sum-capacity for the symmetric $K$-user channel. Interestingly, for the K user channel  having only the $K$-th cognitive transmitter know all other messages is sufficient to achieve capacity, i.e., cognition at transmitters 2 to $K-1$ is not needed. 
 Next, 
 the sum-capacity of the symmetric Gaussian noise channel is characterized to within a constant additive and multiplicative gap. The proposed achievable scheme for the additive gap is based on Dirty Paper Coding and can be thought of as a MIMO-broadcast scheme where only one encoding order is possible due to the message sharing structure. As opposed to other multiuser interference channel models, a single scheme suffices for both the weak and strong interference regimes. With this scheme, the generalized degrees of freedom (gDoF) is shown to be a function of $K$, in contrast to the non cognitive case and the broadcast channel case. 
Interestingly, it is show that as the number of users grows to infinity the gDoF of the $K$-user cognitive interference channel with cumulative message sharing tends to the gDoF of a broadcast channel with a $K$-antenna transmitter and $K$ single-antenna receivers. The analytical additive additive and multiplicative gaps are a function of the number of users. { Numerical evaluations of inner and outer bounds show that the actual gap is less than the analytical one.}
\end{abstract}


\section{Introduction}
\label{sec:intro}


One of the many promising uses of the recently emerged  {\em cognitive radio technology}   has been to  enhancing spectral management by allowing artificially intelligent secondary users (cognitive radios) to exploit the same frequency band without significantly degrading the  performance of licensed/primary users. 
Cognitive radios are capable of  searching for available unused spectrum (interweave), they can operate simultaneously with primary users as long as the interference caused is within an acceptable level (underlay), or it can exploit  knowledge of the messages of primary users   through encoding schemes to cancel interference (overlay)~\cite{goldsmith2009breaking}.

The cognitive radio channel, first introduced in~\cite{devroye_IEEE}, falls into the overlay category, and consists of two source-destination pairs in which one of the transmitters called the {\em secondary transmitter} has non-causal a priori knowledge of the message of the other transmitter known as the {\em primary transmitter}. This non-causal message knowledge idealizes a cognitive radio's ability  to overhear other transmissions and exploit them to either cancel them out at their own receiver or aid in their transmission. 
For the state-of-the-art on the two-user cognitive channel we refer the reader to~\cite{rini:journal1,rini:journal2}. In particular, the capacity of the semi-deterministic two-user cognitive channel is known~\cite{rini:journal1}; the capacity of the Gaussian noise channel is known exactly for most channel parameters, and to within one bit otherwise~\cite{rini:journal2}.

In this paper we extend the two-user cognitive interference channel to the case of $K$ users. 
The $K$-user cognitive interference channel analyzed in this work consists of one primary and $K-1$ secondary,  or cognitive, users. 
We assume a {\em cumulative} message cognition structure introduced in \cite{nagananda2009information} for the three-user channel, and extended here to $K$ users, whereby user 1 is the primary user, and cognitive users~$i$, $i\in[2:K]$, know the messages of user~1 through $i-1$. 
The cumulative message cognition model is inspired by the concept of overlaying, or layering, cognitive networks. In particular, we consider multiple types of devices sharing the spectrum. The first ``layer'' consists of the primary users. Each additional cognitive layer transmits simultaneously with the previous layers (overlay) given the lower layers' codebooks.
This may enable  them to learn the lower layers' messages and use this to aid the lower layers' transmission, or to combat interference at their own receivers. 
For this model, we are interested in the impact the cumulative message cognition has on the sum-capacity, or network throughput, and how it extends known results for the two-user case~\cite{rini:journal2}.
We are also interested in how cumulative message cognition differs from $K$-user channel models such as the $K$-user interference channel (with no cognition) or the $K$-antenna broadcast channel (where every user knows all messages). 


\subsection{Past Work}
The literature on the fundamental performance of multi-user cognitive interference channels is limited, in part due to the fact that the two-user counterpart is not yet fully understood ~\cite{rini:journal1, rini:journal2}. The only other work on a $K$-user cognitive interference channel with $K>3$ is, { to the best of our knowledge}, that of~\cite{vishwanathjafarianmultiusercognitive}. In~\cite{vishwanathjafarianmultiusercognitive} the channel model consists of one primary user and $K-1$ parallel cognitive users; each cognitive user only knows the primary message in addition to their own message (thus not a cumulative message structure); the cognitive users do not cause interfere to one another but only to the primary receiver and are interfered only by the primary transmitter (whereas we consider here a fully connected $K$-user interference channel model); for this channel model the capacity in the ``very strong'' interference regime is obtained by using lattice codes~\cite{vishwanathjafarianmultiusercognitive}. Related as well to $K$-user cognitive channels is the work in \cite{gamalCoMP} where the Degrees of Freedom (DoF) of a $K$-user interference channel ($K$ independent messages) in which each transmitter, in addition to its own message, has access to a subset of the other users' messages, is obtained. We will be interested in characterizing the extension of the DoF -- the Generalized DoF (gDoF), as well as capacity to within a constant gap  -- for one particular message knowledge structure. 

While not much work on $K>3$ channels exists,  in~\cite{nagananda2009information,  nagananda2011, nagananda2010achievable,  mirmohseni2011capacity, Myungstron} different three-user cognitive channels are considered; we note that the models differ from the one considered here either in the number of transmitter/receivers, or in the message sharing/cognition structure in all but ~\cite{nagananda2009information,  nagananda2011}. 
In the more comprehensive~\cite{nagananda2011}, several types of 3-user cognitive interference channels 
are proposed: that with ``cumulative message sharing'' (CMS) as considered here, that with ``primary message sharing'' where the message of the single primary user is known at both cognitive transmitters (who do not know each others' messages), and finally ``cognitive only message sharing'' (CoMS) where there are two primary users who do not know each others' message and a single cognitive user which knows both primary messages.  Achievable rate regions for are obtained which are evaluated in Gaussian noise. The CoMS mechanism yields almost the same message structure as in the interference channel with a cognitive relay  -- identical if the relay were to further have a message of its own (see \cite{Rini:CIFC-CR, cognitiverelaysriram} and references therein for the  interference channel with a cognitive relay). 
In~\cite{nagananda2010achievable} the CoMS was first introduced where an achievable rate region was obtained which employs a combination of superposition coding and Gel'fand-Pinsker's binning and numerically evaluated for the Gaussian noise channel. 
In~\cite{mirmohseni2011capacity} the CoMS structure is assumed and the cognitive user is furthermore assumed not to interfere with the primary users; an inner and an outer bound are obtained.
In~\cite{Myungstron} capacity under ``strong interference'' for the CoMS is obtained. We thus emphasize that the channel considered here is more general than others studied as we consider $K$ users, a fully connected interference channel, and consider the less studied CMS sharing structure. 

%

\subsection{Contributions}


The main contributions of this work are:
\begin{enumerate}

\item We derive a novel and general outer bound region that reduces to the outer bound of~\cite{rini:journal1} for the two-user case. The bound is valid for any memoryless channel and any number of users. 
The bound does not contain any auxiliary random variables and is therefore computable for many channel of interest, including the Gaussian noise channel.

 
\item We determine the sum-capacity the 3-user Linear Deterministic Approximation of the Gaussian noise channel at high-SNR for any channel parameters. This optimal scheme inspires a scheme for the $K$-user symmetric channel. This latter scheme only requires cognition of all messages at one transmitter while all the others can be non-cognitive.


\item We derive the sum-capacity for the symmetric Gaussian Gaussian noise channel with $K$ users to within a constant and a multiplicative gap.  

The additive gap is a function of the number of users and grows as $(K-2)\log_2(K-2)$. The proposed achievable scheme is based on Dirty Paper Coding and can be thought of as a MIMO-broadcast scheme where only one encoding order is possible due to the cumulative message sharing mechanism. As opposed to other multiuser interference channel models, a single scheme suffices for both the weak and strong interference regimes. Moreover no interference alignment of structured coding seems to be needed. {Numerical evaluations show that the actual gap is less than the analytical one; this is so because of necessary crude bounding steps needed to obtain analytically tractable sum-rate expressions.}

The multiplicative gap is $K$ and is achieved by having all users beamform to the primary user.

\item The normalized gDoF, defined as the pre-log of the sum-capacity as a function of SNR normalized by the number of users $K$, is shown to be a function of $K$. This is in contrast with other interference channel models, like the non-cognitive case or the broadcast channel, where the gDoF are the same for any $K$ (except for a discontinuity point). Interestingly, it is show that as the number of users grows to infinity the gDoF of the $K$-user cognitive interference channel with cumulative message sharing tends to the gDoF of a broadcast channel with a $K$-antenna transmitter and $K$ single-antenna receivers. 

\end{enumerate}

\subsection{Paper Organization}
The paper is organized as follows.
Section~\ref{sec:ChModel} describes channel model. 
Section~\ref{sec:outer} contains our novel outer bound region; first the 3-user case is considered to highlight the key `side information' idea and then it is extended to any number of users.
In Section~\ref{sec:LDC} we first derive the sum-capacity of the 3-user Linear Deterministic Approximation of the Gaussian noise channel at high-SNR for any channel parameters and then extend it to the symmetric $K$-user case; we also compare the sum-capacity of the interference channel with cumulative message sharing mechanisms with other interference channel models.
In Section~\ref{sec:AWGN} we derive the sum-capacity of the symmetric Gaussian noise channel to within an { additive and multiplicative gap; we use a DPC-based scheme inspired by MIMO-BC with only encoding order possible due to the cumulative message sharing mechanism and beamforming to the primary user; we further show by numerical optimization of inner and outer bounds that the actual gap is less than the theoretical one.}
The gDoF is also derived and shown to be a function of the number of users; {as for other interference models, the gDoF and the sum-capacity of the  Linear Deterministic Approximation of the Gaussian noise channel at high-SNR coincide.}
Section~\ref{sec:conclusion} concludes the paper.

\section{Channel Model}
\label{sec:ChModel}

\subsection{The General Memoryless Channel}
\label{sec:ChModel:general}

The general memoryless $K$-user cognitive interference channel with cumulative message sharing ($K$-CIFC-CMS) consists of $K$ source-destination pairs sharing the same physical channel with some transmitters having non-causal knowledge of the messages of other transmitters. 
Here  transmitter~1 is referred to as the primary user and is assumed to have no cognitive abilities. Transmitter~$i$, $i\in[2:K]$, is non-causally cognizant of the messages of the users with index smaller than $i$. More formally,  the $K$-CIFC-CMS channel consists of
\begin{itemize}
\item Channel  inputs $X_i \in \mathcal{X}_i, \ i\in[1:K]$, 
\item Channel outputs $Y_i \in \mathcal{Y}_i, \ i\in[1:K]$, 
\item A memoryless channel with joint transition probability (or conditional channel distribution)  $\mathbb{P}(Y_1,\ldots,Y_K|X_1,\ldots,X_K)$,
\item Messages $W_i$ which are known to users $1, 2, \ldots, i$, $i\in [1:K]$.
\end{itemize}
A code with non-negative rate vector $(R_1,\ldots,R_K)$ and blocklength $N$ is defined by
\begin{itemize} 
\item
Messages $W_i, \ i\in[1:K]$, uniformly distributed over $[1:2^{N R_i}]$ and independent of everything else, 
\item Encoding functions $f_i^{(N)} : [1:2^{N R_1}] \times \ldots \times [1:2^{N R_i}] \to \mathcal{X}_i^N$ such that $X_i^N := f_i^{(N)}(W_1,\ldots,W_i)$, $i\in[1:K]$,
\item Decoding functions $g_i^{(N)} : \mathcal{Y}_i^N \to [1:2^{N R_i}]$ such that $\widehat{W}_i = g_i^{(N)}(Y_i^N)$, $i\in[1:K]$, 
\item 
Probability of error $P_{e}^{(N)} := \max_{i\in[1:K]}\mathbb{P}[\widehat{W}_i \not= W_i].$
\end{itemize}
The capacity of the $K$-CIFC-CMS channel consists of all non-negative rate tuples $(R_1,\ldots,R_K)$ for which there exist a sequence of codes indexed by the block length $N$ such that $P_{e}^{(N)} \rightarrow$ 0 as $N \rightarrow \infty$. Since the decoders cannot cooperate and the channel is used without feedback, the capacity may be shown to depend only on the marginal noise distributions rather than the joint noise distribution by an argument similar to that used for the Broadcast Channel (BC)~\cite{coverBC}.

In this work we shall focus on the following two channel models.

\subsection{The Gaussian Noise Channel}
\label{sec:ChModel:awgn}

\begin{figure}
\centering
\includegraphics[width=10cm]{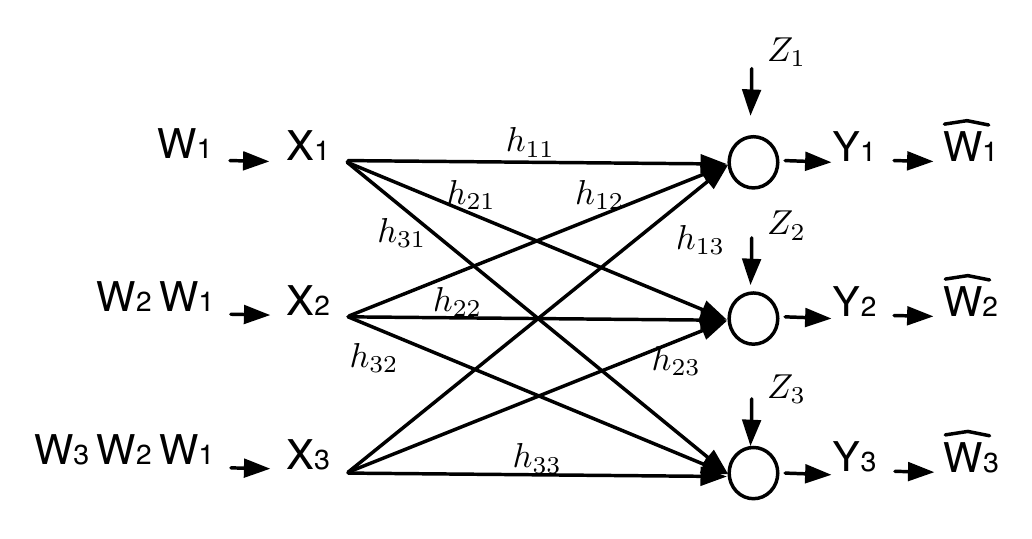}
\caption{The Gaussian $3$-CIFC-CMS.}
\label{fig:Gaussian 3-CIFC-CMS Channel model}
\end{figure}

The single-antenna complex-valued $K$-CIFC-CMS with Additive White Gaussian Noise (AWGN), shown in Fig.~\ref{fig:Gaussian 3-CIFC-CMS Channel model} for the case $K=3$, has input-output relationship
\begin{subequations}
\begin{align}
Y_\ell = \sum_{i\in[1:K]} h_{\ell i}X_i+ Z_\ell, \ \ell\in [1:K], 
\end{align}
where, without loss of generality, the inputs are subject to the power constraint
\begin{align}
\mathbb{E}[|X_i|^2] \leq 1, \ i\in [1:K], 
\end{align}
and the noises are marginally proper-complex Gaussian random variables with parameters
\begin{align}
Z_\ell\sim \mathcal{N}(0,1),\  \ell\in [1:K]. 
\end{align}
\label{eq:AWGN channel model}
\end{subequations}
The channel gains $h_{ij}$, $(i,j)\in[1:K]^2$, are constant and therefore known to all terminals. Without loss of generality we may assume the direct links $h_{ii}$, $i\in [1:K]$ to be real-valued and non-negative since the receiver~$i$ can always compensate for the phase of one channel gain. 

The Generalized Degrees-of-Freedom (gDoF) of the {\em symmetric} Gaussian channel is a performance metric that characterizes the high-SNR behavior of the sum-capacity and is defined as follows.
Let $\mathsf{SNR}$ be a non-negative number and parameterize
\begin{subequations}
\begin{align}
|h_{i i}|^2    &:= \mathsf{SNR}, \ i\in [1:K], \\
|h_{\ell i}|^2 &:= \mathsf{SNR}^{\alpha}, \ (\ell,i)\in[1:K]^2, \ell\not= i,
\end{align}
\label{eq:ch parm}
\end{subequations}
for some non-negative $\alpha$.  The gDoF is
\begin{align}
d(\alpha) &:= \lim_{\mathsf{SNR}\to+\infty} \frac{C_{\Sigma}}{\log(1+\mathsf{SNR})},
\label{eq:def gdof}
\end{align}
where $C_{\Sigma}:=\max\{R_1+\ldots+R_K\}$ and where the maximization is over all achievable rates.
The sum-capacity is said to be known to within a constant gap of $\mathsf{b}$~bits if one can show rates $R_{\Sigma}^{\rm(in)}$ and $R_{\Sigma}^{\rm(out)}$ such that
\begin{align}
R_{\Sigma}^{\rm(in)}\leq C_{\Sigma}\leq R_{\Sigma}^{\rm(out)}
\leq R_{\Sigma}^{\rm(in)} + \mathsf{b}\log(2).
\label{eq:def cap within a gap}
\end{align}
The gDoF and constant gap characterization of the symmetric sum capacity imply that
\[
C_{\Sigma} = d(\alpha) \log(1+\mathsf{SNR}) + o(1),
\]
where $o(1)$ indicates a quantity that is finite at all $\mathsf{SNR}$.

\subsection{Linear Deterministic Approximation of the Gaussian Noise Channel}
\label{sec:ChModel:ldc}

The Linear Deterministic approximation of the Gaussian Noise Channel at high SNR (LDC) was first introduced in~\cite{avestimehr_diggavi_tse_det} to allow focussing on the signal interactions rather than on the additive noise. The proposed framework has been very powerful in revealing key issues for the problem of communicating over interfering networks and the insights gained for the LDC have often been translated into capacity results to within a constant gap for any finite SNR~\cite{etkin_tse_wang, rini:journal2, suh2010feedback}. In light of these success stories we also start our investigation from the LDC.
The LDC has input-output relationship given by
\begin{align}
Y_\ell = \sum_{i\in[1:K]} \mathbf{S}^{m-n_{\ell i}} X_i,  \ \ell\in [1:K], 
\label{eq:LDC channel}
\end{align}
where $m := \max\{n_{ij}\}$, $\mathbf{S}$ is the binary shift matrix of dimension $m$,
all inputs and outputs are binary column vectors of dimension $m$,
the summation is bit-wise over of the binary field, and the channel gains $n_{\ell i}$ for $(\ell,i)\in[1:K]^2,$ are positive integers. In a {\em symmetric} LDC all direct links have the same strength $n_{i i} = n_{\rm d} \geq 0, i\in [1:K],$ and all the interfering links have the same strength $n_{\ell i} = n_{\rm i} = \alpha \ n_{\rm d} \geq 0, (\ell,i)\in [1:K]^2, \ \ell\not=i$. Note that the subscript ${\rm i}$ (roman font) of $n_{\rm i}$ stands for `interference' and is not an index; as such it should not be confused with index $i$ (italic font).

The channel in~\eqref{eq:LDC channel} can be thought of as the high SNR approximation of the channel in~\eqref{eq:AWGN channel model} with their parameters related as $n_{ij} = \lfloor \log(1+|h_{ij}|^2 ) \rfloor, \ (i,j)\in[1:K]^2$. 

\section{Outer Bound}
\label{sec:outer}

In this section we derive an outer-bound region for the general memoryless $K$-CIFC-CMS as defined in Section~\ref{sec:ChModel:general}. We start with the case of $K=3$ users to highlight the main proof techniques and ease the reader into the extension to any number of users $K\in\mathbb{N}^+$.

\begin{thm}
\label{thm:outer K=3}
The capacity region of the general memoryless $3$-CIFC-CMS is contained in the region defined by
\begin{subequations}
\begin{align}
          R_1 &\leq I(Y_1; X_1,X_2,X_3), \label{eq:K=3 1}
\\        R_2 &\leq I(Y_2;     X_2,X_3| X_1), \label{eq:K=3 2}
\\        R_3 &\leq I(Y_3;         X_3| X_1,X_2), \label{eq:K=3 3}
\\    R_2+R_3 &\leq I(Y_2;     X_2,X_3| X_1)
                  + I(Y_3;         X_3| X_1,X_2, Y_2), \label{eq:K=3 2,3}
\\R_1+R_2+R_3 &\leq I(Y_1; X_1,X_2,X_3)
                  + I(Y_2;     X_2,X_3| X_1, Y_1) \nonumber
\\&               + I(Y_3;         X_3| X_1,Y_1, X_2,Y_2), \label{eq:K=3 1,2,3}
\end{align}
\label{eq:K=3 outer}
\end{subequations}
for some input distribution $P_{X_1,X_2,X_3}$.
The joint conditional distribution $P_{Y_1,Y_2,Y_3|X_1,X_2,X_3}$ can be chosen
so as to tighten the different bounds as long as the conditional
marginal distributions $P_{Y_i|X_1,X_2,X_3}$, $i\in[1:3]$, are preserved.
\end{thm}
\begin{IEEEproof}
The proof can be found in Appendix~\ref{app:proof of thm:outer K=3}.
\end{IEEEproof}

Remarks:
\begin{enumerate}
\item
The region in Th.~\ref{thm:outer K=3} reduces to the outer bound in~\cite[Th. 6]{rini:journal1} by setting $X_3=Y_3=\emptyset$.
\item
The outer bound region in~\eqref{eq:K=3 outer} does contain auxiliary random variables. Moreover, every mutual information term contains all the inputs. These two facts imply that the outer bound region in Th.~\ref{thm:outer K=3} can be easily evaluated for many channel of interest. For example, for the Gaussian noise channel in Section~\ref{sec:ChModel:awgn}, the ``Gaussian maximizes entropy'' principle suffices to show that jointly Gaussian inputs exhaust the outer bound.
\item
The sum-capacity bound in~\eqref{eq:K=3 1,2,3} is obtained by giving $S_i$ as side information to receiver~$i$, $i\in[1:K]$, where $S_i = [S_{i-1}, W_{i-1}, Y_{i-1}^N]$ starting with $S_1=\emptyset$. With this ``nested'' side information, the mutual information terms can be expressed in terms of entropies which may be recombined in ways that can be easily single-letterized. This form of the side information allows us to extend the result from the 3-user case to any number of users.
\item
The mutual information terms in~\eqref{eq:K=3 1,2,3} have the form $I(Y_i; X_{i},\ldots,X_{K}|X_{1},Y_{1},\ldots,X_{i-1},Y_{i-1})$, $1 \leq i \leq K$, which can be given the following interpretation. 
Since message $W_i$ is available at transmitters $i$ through $K$, inputs $(X_{i},\ldots,X_{K})$ are ``informative'' for receiver~$i$, while inputs $(X_{1},\ldots,X_{i-1})$ are independent of $W_i$; receiver~$i$ decodes from $Y_{i}$ the information carried in $(X_{i},\ldots,X_{K})$ that could not be recovered by users with lesser index as represented by $(X_{1},Y_{1},\ldots,X_{i-1},Y_{i-1})$.
\end{enumerate}

Th.~\ref{thm:outer K=3} can be extended to a general memoryless $K$-CIFC-CMS.
\begin{thm}
\label{thm:outer K}
The capacity region of the general memoryless $K$-CIFC-CMS is contained in the region defined by
\begin{subequations}
\begin{align}
                  R_i &\leq I(Y_i; X_i,\ldots,X_K| X_1,\ldots, X_{i-1}), \ i \in [1:K],
\label{eq:outer bound general K single}
\\ \sum_{i=j}^{K} R_j &\leq \sum_{i=j}^{K} I(Y_j; X_j,\ldots,X_K| X_1,\ldots, X_{j-1},Y_i,\ldots,Y_{j-1}),  \ i \in [1:K],
\label{eq:outer bound general K progressive}
\end{align}
\label{eq:K outer}
\end{subequations}
for some input distribution $P_{X_1,\ldots,X_K}$.
Moreover, each rate bound in~\eqref{eq:outer bound general K progressive} may be tightened with respect to the channel conditional distribution as long as the channel conditional marginal distributions are preserved.
\end{thm}
\begin{IEEEproof}
The proof is found in Appendix~\ref{app:proof of thm:outer K}.
\end{IEEEproof}

In the following section we shall derive achievable schemes matching the sum-capacity outer bound in Th.~\ref{thm:outer K} for the LDC in~\eqref{eq:LDC channel} and schemes that achieve the sum-capacity outer bound to within a constant bounded gap regardless of the channel parameters for the Gaussian channel in~\eqref{eq:AWGN channel model}.

\section{Sum-capacity for the Linear Deterministic $K$-CIFC-CMS}
\label{sec:LDC}

In Sections~\ref{sec:3-LDC-gen-out} and~\ref{sec:3-LDC-gen-in} we determine the sum-capacity of the LDC with $K=3$ users and any value of the channel gains. 
In Sections~\ref{sec:K-LDC-sym-out} and~\ref{sec:K-LDC-sym-in} we derive the sum-capacity for any $K$ but for symmetric channel gains only.
The main results of this section are
\begin{thm}
\label{thm:3-LDC-all}
The sum-capacity bound in~\eqref{eq:K=3 1,2,3} is achievable for the LDC $3$-CIFC-CMS with generic channel gains.
\end{thm}
\begin{thm}
\label{thm:K-LDC-sym}
The sum-capacity bound in~\eqref{eq:outer bound general K progressive} is achievable for the LDC $K$-CIFC-CMS with symmetric channel gains. { The capacity achieving scheme only requires cognition of all messages at one single transmitter.}
\end{thm}
The rest of the section is devoted to their proofs.

\subsection{Sum-capacity upper bound for the 3-user case and generic channel gains}
\label{sec:3-LDC-gen-out}

The sum-capacity outer bound in Th.~\ref{thm:outer K=3} specialized to a deterministic 3-CIFC-CMS (i.e., $H(Y_i|X_1,X_2,X_3)=0, i\in[1:3]$) gives the following sum-capacity upper-bound
\begin{align*}
  R_1+R_2+R_3 \leq \max \Big\{ H(Y_1)+ H(Y_2| X_1,Y_1) + H(Y_3| X_1,Y_1,X_2,Y_2)\Big\},
\end{align*}
where the maximization is over all possible joint distributions $P_{X_1,X_2,X_3}$.
For the LDC in~\eqref{eq:LDC channel} with $K=3$ we obtain
\begin{subequations}
\begin{align}
R_1+R_2+R_3  
  &\leq \max\{n_{11},n_{12},n_{13}\}
\\&+f(n_{22},n_{23}|n_{12},n_{13}) \label{eq:f fun}
\\&+[n_{33}-\max\{n_{13},n_{23}\}]^+, 
\label{eq:LDC K=3 sum-capacity upper 2}
\end{align}
\label{eq:LDC K=3 sum-capacity upper}
\end{subequations}
where $f(c,d|a,b)$ in~\eqref{eq:f fun} follows from~\cite[eq.(5)]{prabhakaran2011interference} and is defined as
\begin{align*}
  &f(c,d|a,b) :=
\left\{\begin{array}{l l}
  \max\{c+b,a+d\}-\max\{a,b\} &\ \text{if $c-d\not=a-b$},\\
  \max\{a,b,c,d\}-\max\{a,b\} &\ \text{if $c-d=a-b$}.\\
\end{array}\right.
\end{align*}
The bound in~\eqref{eq:LDC K=3 sum-capacity upper} follows by maximizing each mutual information term individually as
\begin{align*}
H(Y_1)
  &= H(\mathbf{S}^{m-n_{11}} X_1+\mathbf{S}^{m-n_{12}} X_2+\mathbf{S}^{m-n_{13}} X_3) 
\\&\leq \max\{n_{11},n_{12},n_{13}\}, 
\\
H(Y_2| X_1,Y_1)
&=H(\mathbf{S}^{m-n_{22}}X_2+\mathbf{S}^{m-n_{23}}X_3| X_1, \mathbf{S}^{m-n_{12}}X_2+\mathbf{S}^{m-n_{13}}X_3)
\\&\leq H(\mathbf{S}^{m-n_{22}}X_2+\mathbf{S}^{m-n_{23}}X_3| \mathbf{S}^{m-n_{12}}X_2+\mathbf{S}^{m-n_{13}}X_3)
\\&\leq  f(n_{22},n_{23}|n_{12},n_{13}),
\\
H(Y_3| X_1,Y_1,X_2,Y_2)
&=H(\mathbf{S}^{m-n_{33}} X_3| X_1,X_2,\mathbf{S}^{m-n_{13}} X_3,\mathbf{S}^{m-n_{23}} X_3)
\\&\leq H(\mathbf{S}^{m-n_{33}} X_3| \mathbf{S}^{m-\max\{n_{13}, n_{23}\}} X_3)
\\&\leq [n_{33}-\max\{n_{13}, n_{23}\}]^+,
\end{align*}
where $[x]^+ := \max\{0,x\}$.
Notice that i.i.d. Bernoulli(1/2) input bits simultaneously maximize each of the above entropy terms.

\subsection{Achievability of the sum-capacity upper bound for the 3-user case and generic channel gains}
\label{sec:3-LDC-gen-in}

In the following, depending on whether $[n_{33}-\max\{n_{13}, n_{23}\}]^+$ in~\eqref{eq:LDC K=3 sum-capacity upper 2} is zero or positive, different interference scenarios are identified and transmission schemes that are capable of achieving the sum-capacity upper bound in~\eqref{eq:LDC K=3 sum-capacity upper} are proposed.
In particular:

{\bf Case~1:}
If the signal sent by the most cognitive transmitter is received the weakest at the intended destination, that is, if 
\begin{align}
n_{33}\leq \max\{n_{13},n_{23}\},
\label{eq:implies r3=0}
\end{align}
the sum-capacity in~\eqref{eq:LDC K=3 sum-capacity upper} becomes 
\begin{align*}
R_1+R_2+R_3  
\leq  \max\{n_{11},n_{12},n_{13}\}+f(n_{22},n_{23}|n_{12},n_{13}).
\end{align*}
The condition in~\eqref{eq:implies r3=0} corresponds to the case $H(Y_3| X_1,Y_1,X_2,Y_2)=0$, i.e., conditioned on $(X_1,X_2)$ the signal received at the ``most cognitive'' receiver is a degraded version of the signal received at the other two receivers. Recall that user~3 can send information to all receivers as it knows all messages. The condition in~\eqref{eq:implies r3=0} implies that the signal $X_3$ can convey more information to receivers~1 and~2 that it can to the intended receiver~3. In this case, one might thus suspect that $R_3=0$ is optimal and that the best use of the cognitive capabilities of user~3 is to ``broadcast'' to the non-intended receivers. We will next show that this is indeed the case.

We set $R_3=0$ and we therefore convert the LCD $3$-CIFC-CMS into a deterministic $2$-CIFC-CMS where user~1 is the primary user (with input $X_1$ and output $Y_1$) and the cognitive user has vector input $[X_2,X_3]$ and output $Y_2$. The capacity of a general deterministic 2-user cognitive interference channel is~\cite[Th. 12]{rini:journal1} 
\begin{align*}
 R_1 &\leq H(Y_1), \    R_2 \leq H(Y_2| X_1),
\\R_1+R_2 &\leq H(Y_1)+ H(Y_2| X_1, Y_1)
\end{align*}
for some input distribution $P_{X_1,[X_2,X_3]}$.
Hence the sum-capacity is
\begin{align*}
 R_1+R_2
   &= \max_{P_{X_1,[X_2,X_3]}}\Big\{ H(Y_1)+ H(Y_2| X_1, Y_1) \Big\}
 \\&= \max\{n_{11},n_{12},n_{13}\}+f(n_{22},n_{23}|n_{12},n_{13}),
\end{align*}
which proves our claim.

\medskip
{\bf Case~2:}
In the regime not covered by the condition in~\eqref{eq:implies r3=0}, that is, for
\begin{align}
n_{33}> \max\{n_{13},n_{23}\},
\label{eq:implies r3>0}
\end{align}
the sum-capacity in~\eqref{eq:LDC K=3 sum-capacity upper} becomes 
\begin{align*}
R_1+R_2+R_3  
\leq  \max\{n_{11},n_{12},n_{13}\}+f(n_{22},n_{23}|n_{12},n_{13}) + n_{33}-\max\{n_{13}, n_{23}\}.
\end{align*}
Is this case, the condition in~\eqref{eq:implies r3>0} suggests that the intended signal at receiver~3 is sufficiently strong to be able to support a non-zero rate. The form of the sum-capacity also suggests that a plausible strategy is to use the optimal strategy for Case~1 and ``sneak in'' extra bits for user~3 in such a way that they do not appear at the other receivers. We next show that this is optimal.

We split the signal of transmitter~3 in two parts
\begin{align*}
 X_3 := X_{3a} + X_{3b},
\end{align*}
where $X_{3a}$ is intended to mimic the scheme for Case~1 (i.e., as if user~2 had input $[X_2,X_{3a}]$)
and $X_{3b}$ carries the information to $Y_3$, possibly ``pre-coded'' against the interference of $(X_1,X_2,X_{3a})$,
and such that $X_{3b}$ is not received at receivers~2 and~3. We define
\begin{align*}
 X_{3b} := S^{\max\{n_{13},n_{23}\} } V_3, 
\end{align*}
for some vector $V_3$ defined in the following.
Note that the shift caused by $S^{\max\{n_{13},n_{23}\} }$  is such that $V_3$ is not received at $Y_1$ and at $Y_2$. We note that $V_3$ is ``private information'' for receiver~3 that is dirty paper coded against the interference caused by $[X_1,X_2,X_{3a}]$ at receiver~3; with this receiver~3 is virtually interference-free.
We then implement the optimal strategy for Case~1 with $[X_1,X_2,X_{3a}]$ and with the remaining bits in $X_{3b}$ we transmit to receiver~3 thereby achieving the sum-capacity in~\eqref{eq:LDC K=3 sum-capacity upper}.

\subsection{Example of sum-capacity optimal schemes for the 3-user case and symmetric channel gains}
\label{sec:3-LDC-sym-example}

We present here some concrete examples of the achievability scheme presented in Section~\ref{sec:3-LDC-gen-in}.

We consider first the symmetric scenario with $n_{\rm d} >0, \ n_{\rm i} = n_{\rm d} \ \alpha, \ \alpha\geq0$. Define the normalized sum-capacity as
\begin{align*}
d_{\Sigma}(\alpha; K) := \frac{\max\{R_1+R_2+R_3\}}{n_{\rm d}}.
\end{align*}
Note that when $n_{\rm d}=0$ the channel reduces to a broadcast channel from transmitter $[X_2,X_3]$ to receivers $Y_1$ and $Y_2$ (receiver~3 cannot be reached by its transmitter and hence $R_3=0$ is optimal; similarly the primary user cannot reach its intended destination and cannot deliver any information to the other destinations, hence $X_1=0$ is optimal); the capacity region of a deterministic broadcast channel is known~\cite{martonBC} and for the symmetric LDC with $n_{\rm d}=0$ it reduces to $R_1+R_2 = 2 n_{\rm i}$.
 
\begin{figure}
\centering
\includegraphics[width=8cm]{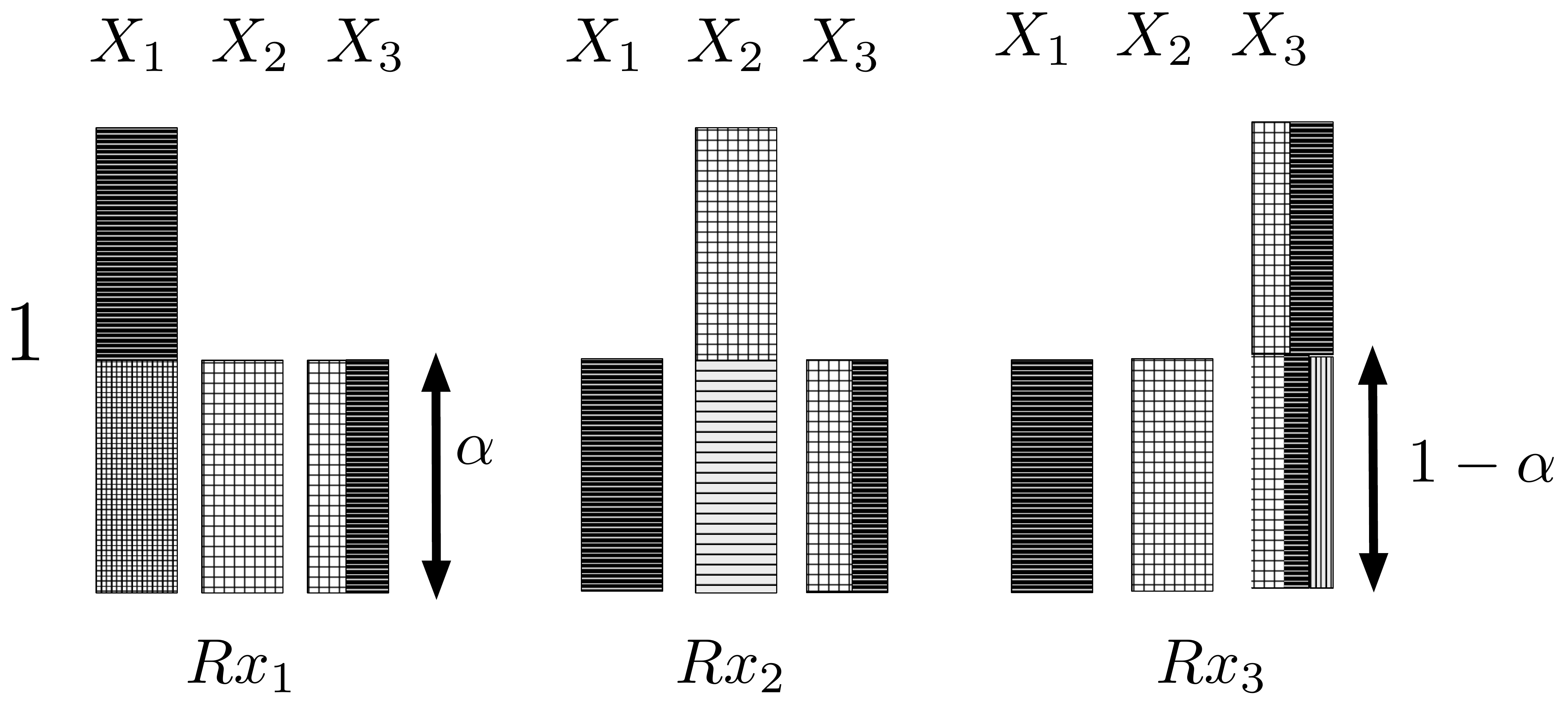}%
\caption{LDC 3-CIFC-CMS in weak interference with $\alpha =1/2$.
The achievable rates are $R_1/n=R_2/n=1, R_3/n=1-\alpha$ thereby achieving the sum-capacity upper bound in~\eqref{eq:LDC K=3 sum-capacity upper} under the condition in~\eqref{eq:implies r3>0}.}
\label{fig:weak}
\end{figure}

When $n_{\rm d}>0$ the sum-capacity can be expressed as
\begin{align*}
d_{\Sigma}(\alpha; 3) 
  &= \max\{1,\alpha\} + \frac{f(n_{\rm d},n_{\rm d}\ \alpha;n_{\rm d}\ \alpha,n_{\rm d}\ \alpha)}{n_{\rm d}} + [1-\alpha]^+
\\&= \left\{\begin{array}{ll}
 3\max\{1,\alpha\}-\alpha & \text{ for $\alpha\not=1$}, \\
 1 & \text{ for $\alpha=1$}. \\
\end{array}\right.
\end{align*}
Fig~\ref{fig:weak} shows an example of the achievable strategy for {\em weak interference} defined as $\alpha<1$ (corresponding to Case~2 in Section~\ref{sec:3-LDC-gen-out}). The case $\alpha=1$ corresponds to a channel where all received signals are statistically equivalent and therefore its capacity region is as for the 3-user Multiple Access  Channel. The {\em strong interference} regime defined as $\alpha>1$ (corresponding to Case~1 in Section~\ref{sec:3-LDC-gen-out}) is not explicitly considered as the achievable strategy is the same as for the weak interference regime except for the fact that the most cognitive user does not send any message for himself as its bits would create interference at the non-intended receivers.
Notice the important role of cognition in Fig.~\ref{fig:weak}.
The third transmitter (cognitive of all 3 messages) sends a linear combination of the messages of users~1 and~2 in such a way that the effect of the aggregate interference is neutralized at all receivers. This leaves the receivers of users~1 and~2 interference-free. The third transmitters also sends some ``private'' information bits in such a way that these bits do not appear at the other receivers. It is important also to observe that user~2, who is cognizant of the message of user~1, does not use the knowledge in the encoding process. In other words, user~2 need not be cognizant in order to achieve the sum-capacity in the symmetric case.


%

\subsection{Sum-capacity upper bound for the $K$-user case and symmetric channel gains}
\label{sec:K-LDC-sym-out}

For the $K$-user symmetric LDC the sum-capacity is upper bounded by
\begin{align}
 &\frac{\sum_{k=1}^{K} R_k}{n_{\rm d}}
 \leq 
\left\{\begin{array}{ll}
  K\max\{1,\alpha\}-\alpha & \text{ for $\alpha\not=1$}, \\
  1 & \text{ for $\alpha=1$}, \\
\end{array}\right.
\label{eq:LDC K sum-capacity upper}
\end{align}

The proof that the sum-capacity upper bound in Th.~\ref{thm:outer K} evaluates to the expression in~\eqref{eq:LDC K sum-capacity upper} is provided next. For the $K$-user symmetric LDC with $m=n_{\rm d} \max\{1,\alpha\}$ the sum-capacity is upper bounded by
\begin{align*}
\sum_{k=1}^{K} R_k
  &\leq \sum_{k=1}^{K} 
  H\left(Y_k| X_1,\ldots,X_{k-1}, Y_1,\ldots,Y_{k-1}\right)
\\&= \sum_{k=1}^{K-1} 
H\left(\mathbf{S}^{m-n_{\rm d}}X_k+\mathbf{S}^{m-n_{\rm i}}\left(\sum_{i=k+1}^{K} X_i\right)
\Big| X_1,\ldots,X_{k-1}, \mathbf{S}^{m-n_{\rm i}}\left(\sum_{i=k}^{K} X_i\right) \right)
\\&+H\Big(\mathbf{S}^{m-n_{\rm d}}X_K|X_1,\ldots,X_{K-1},\mathbf{S}^{m-n_{\rm i}}X_K\Big)
\\&\leq 
 \sum_{k=1}^{K-1} H\Big( (\mathbf{S}^{m-n_{\rm d}}+\mathbf{S}^{m-n_{\rm i}})X_k\Big)
 +H\Big(\mathbf{S}^{m-n_{\rm d}}X_K|\mathbf{S}^{m-n_{\rm i}}X_K\Big)
\\&\leq 
(K-1)  \max\{n_{\rm d},n_{\rm i}\} + [n_{\rm d}-n_{\rm i}]^+
\\&= n_{\rm d} \Big( K\max\{1,\alpha\}-\alpha\Big).
\end{align*}
The discontinuity at $\alpha=1$ in~\eqref{eq:LDC K sum-capacity upper} is because when $n_{\rm d}=n_{\rm i}$ all received signal are equivalent, i.e.,  $Y_1=\ldots=Y_K=\sum_{i=1}^{K}X_i$, and the channel reduces to a $K$-user MAC with sum-capacity $\max H(Y_1)= n_{\rm d}$.

\subsection{Achievability of the sum-capacity upper bound for the $K$-user case and symmetric channel gains}
\label{sec:K-LDC-sym-in}

The schemes which were shown to be optimal for LCD $3$-CIFC-CMS in Section~\ref{sec:3-LDC-sym-example} can be extended to any arbitrary number of users. Let $U_j$ $j\in[1:K]$, be the signal intended for receiver~$j$, that is, $U_j$ is only a function of message $W_j$, and composed of i.i.d. Bernoulli(1/2) bits.
Let the transmit signals be
\begin{align*}
   X_j &= U_j, \ j\in[1:K-1],
\\ X_K &= 
\begin{bmatrix}
I_{n_{\rm i}} & 0_{n_{\rm i} \times [n_{\rm d}-n_{\rm i}]^+} \\
0_{[n_{\rm d}-n_{\rm i}]^+ \times n_{\rm i} } & 0_{[n_{\rm d}-n_{\rm i}]^+ \times [n_{\rm d}-n_{\rm i}]^+} \\
\end{bmatrix}
\left(\sum_{j=1}^{K-1} U_j\right)
+
\begin{bmatrix}
0_{n_{\rm i} \times n_{\rm i}} & 0_{n_{\rm i} \times [n_{\rm d}-n_{\rm i}]^+} \\
0_{[n_{\rm d}-n_{\rm i}]^+ \times n_{\rm i} } & I_{[n_{\rm d}-n_{\rm i}]^+} \\
\end{bmatrix}
U_K,
\end{align*}
so that
\begin{align*}
\sum_{j=1}^{K}X_j
=
\begin{bmatrix}
0_{n_{\rm i} \times n_{\rm i}} & 0_{n_{\rm i} \times [n_{\rm d}-n_{\rm i}]^+} \\
0_{[n_{\rm d}-n_{\rm i}]^+ \times n_{\rm i} } & I_{[n_{\rm d}-n_{\rm i}]^+} \\
\end{bmatrix}
\left(\sum_{j=1}^{K} U_j\right).
\end{align*}
where $0_{n \times m} $ indicates the all zero matrix of dimension $n \times m$ and $I_{n}$ the identity matrix of dimension $n$.
With these choices, the signal at receiver~$\ell$, $\ell\in[1:K]$, is
\begin{align*}
  Y_\ell &=S^{m-n_{\rm d}} X_\ell + S^{m-n_{\rm i}}\sum_{j\in[1:K], j\not=\ell}X_j
  \\&= (S^{m-n_{\rm d}}+S^{m-n_{\rm i}})X_\ell +S^{m-n_{\rm i}}\left(\sum_{j=1}^{K}X_j\right)
  \\&= (S^{m-n_{\rm d}}+S^{m-n_{\rm i}})X_\ell, \quad m=\max\{n_{\rm d},n_{\rm i}\}.
\end{align*}
Since the matrix $S^{m-n_{\rm d}}+S^{m-n_{\rm i}}$ is full rank for $n_{\rm d}\not=n_{\rm i}$, receiver~$\ell$, $\ell\in[1:K]$, decodes $U_\ell$ from $(S^{m-n_{\rm d}}+S^{m-n_{\rm i}})^{-1}Y_\ell = X_\ell.$ Hence receiver~$\ell$, $\ell\in[1:K-1]$, can decode $m=\max\{n_{\rm d},n_{\rm i}\}$ bits since $X_\ell = U_\ell$, while receiver~$K$ can decode the lower $[n_{\rm d}-n_{\rm i}]^+$ bits of $U_K$ from $X_K$.
Interestingly, receivers from $1$ to $K-1$ are interference free, while receiver~$K$ decodes $n_{\rm i}$ bits of the `interference function' $\sum_{j=1}^{K-1}U_j$.  Notice that cognition is only needed at one transmitter in all interference regimes. This implies that this sum-capacity result holds for all cognitive channels where user $i$ is cognizant of any subset (including the empty set) of the messages of users with index less than $i$.


\subsection{Comparison between different channel models}
\label{sec:comparison}

We compare the symmetric sum-capacity of channels with different levels of cognition. 
Our base line for comparison is the classical $K$-user interference channel without any cognition, whose sum-capacity is~\cite{jafar_gdofK}
\begin{align}
d_{\Sigma}^{\rm(IFC)}(\alpha; K) =
\frac{K}{2}  d_{\Sigma}^{\rm(IFC)}(\alpha; 2) 
\end{align}
and where $d_{\Sigma}^{\rm(IFC)}(\alpha; 2)$ is the so-called W-curve of~\cite{etkin_tse_wang} except for a discontinuity at $\alpha=1$ where $d_{\Sigma}^{\rm(IFC)}(\alpha; K)=1$ for all $K$~\cite{jafar_gdofK}.
Note that, except at $\alpha=1$, the normalized sum-capacity $\frac{1}{K}d_{\Sigma}^{\rm(IFC)}(\alpha; K)$ does not depend on $K$.

At the other end of the spectrum we have the case where all users are cognitive of all messages. In this case the channel is equivalent to a MIMO-BC with $K$ transmit antennas and $K$ single-antenna receivers. Since the system has enough degrees of freedom to zero-force the interference we have
\begin{align}
d_{\Sigma}^{\rm(BC)}(\alpha; K) = K \max\{1,\alpha\},
\end{align}
except for a discontinuity at $\alpha=1$ where $d_{\Sigma}^{\rm(BC)}(\alpha; K)=1$, since in this case all the receivers are statistically equivalent and Time Division Multiple Access (TDMA, or time-sharing) is optimal. Also in this case, except at $\alpha=1$, the normalized sum-capacity $\frac{1}{K}d_{\Sigma}^{\rm(BC)}(\alpha; K)$ does not depend on $K$.

The sum-capacity of the symmetric LDC $K$-CIFC-CMS is given by~\eqref{eq:LDC K sum-capacity upper}, which 
is a function of $K$ after normalization by $K$, i.e.,
\begin{align}
\frac{1}{K}d_{\Sigma}^{\rm(CIFC-CMS)}(\alpha; K) 
=  \max\{1,\alpha\}-\frac{\alpha}{K} 
=  \max\left\{ 1-\frac{\alpha}{K}, \frac{K-1}{K}\alpha \right\}. 
\end{align}
This has the interesting interpretation that CMS looses $\alpha/K$ with respect to $d_{\Sigma}^{\rm(BC)}(\alpha; K)/K$. In other words, as the number of cognitive users increases the CMS sum-capacity approaches the sum-capacity of a fully coordinated broadcast channel.

\begin{figure}
\centering
\includegraphics[width=12cm]{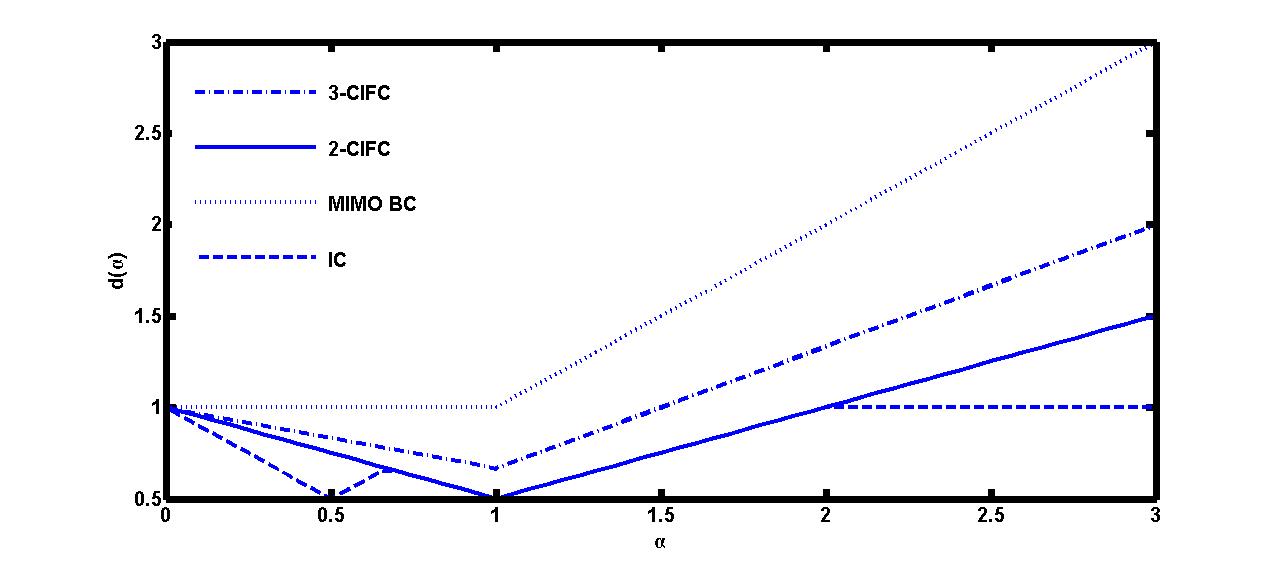}
\caption{$d_{\Sigma}(\alpha; K)/K$ for different channel models. The discontinuity at $\alpha=1$ is not shown where the value is $\frac{1}{K}$.}
\label{fig:comp dnorm}
\end{figure}

Fig.~\ref{fig:comp dnorm} shows the sum-capacity normalized by the number of users for different channel models {(here we do not show the discontinuity at $\alpha=1$)}. We note the increase in performance in all interference regimes when compared to that of 2-user CIFC-CMS and the classical $K$-user interference channel, but a loss with respect to the $K$-user broadcast channel (BC) with K transmit antennas and K single antenna receivers.

\section{Sum-Capacity for Gaussian $K$-CIFC-CMS to within a constant gap}
\label{sec:AWGN}

In this section we derive the sum-capacity for the symmetric Gaussian channel with an arbitrary number of users to within a constant gap. 
For notational convenience we denote the direct link gains as $|h_{\rm d}|$, which can be taken to be real-valued and non-negative without loss of generality, and the interference link gains as $h_{\rm i}$, so that the channel in~\eqref{eq:AWGN channel model} can be rewritten as
\begin{align*}
Y_\ell 
= \Big(|h_{\rm d}|-h_{\rm i}\Big) X_\ell  + h_{\rm i} \Big(\sum_{j=1}^{K}X_j\Big) + Z_\ell, \ \ell\in[1:K].
\end{align*}
The main result sof this section are
\begin{thm}
\label{thm:K-AWGN-dgdof}
The generalized Degrees-of-Freedom of the symmetric $K$-user Gaussian noise channel are
\[
d(\alpha) = K \max\{1,\alpha\} - \alpha
\] 
with a discontinuity at $\alpha=1$ in the special case where all channel gains are the same (in modulo and phase), in which case $d(1)=1$.
\end{thm}
\begin{thm}
\label{thm:K-AWGN-sym-add}
The sum-capacity bound in~\eqref{eq:outer bound general K progressive} is achievable for the symmetric Gaussian $K$-CIFC-CMS to within $6$~bits per channel use for $K=3$ and to within $(K-2)\log_2(K-2) + 3.88$~bits per channel use for $K\geq 4$. 
\end{thm}
\begin{thm}
\label{thm:K-AWGN-sym-mul}
The sum-capacity bound in~\eqref{eq:outer bound general K progressive} is achievable to within a factor $K$ by beamforming to the primary user.
\end{thm}

\subsection{Sum-capacity upper bound for the $K$-user case and symmetric channel gains}
\label{sec:K-AWGN-sym-out}

For the $K$-user symmetric Gaussian channel with $|h_{\rm d}| \not= h_{\rm i}$ the bound in~\eqref{eq:outer bound general K progressive} can be further bounded as (although we can tighten the bound by choosing the `worst noise covariance matrix', we shall use here independent noises)
\begin{align*}
\sum_{k=1}^{K} R_k 
  &\leq \sum_{u=1}^{K} I\Big( X_u,\cdots, X_K; Y_u \Big| X_1,Y_1,\cdots, X_{u-1},Y_{u-1}\Big)
\\&= 
    I\Big( X_1,\cdots, X_K; |h_{\rm d}| X_1  + h_{\rm i} \sum_{i=2}^{K}X_i + Z_1 \Big) 
\\&+ \sum_{u=2}^{K-1} I\Big( X_u,\cdots, X_K; |h_{\rm d}| X_u  + h_{\rm i} \sum_{i=u+1}^{K}X_i + Z_u \Big|
  X_\ell, \ h_{\rm i} \sum_{i=u}^{K}X_i + Z_\ell, \ \ell\in[1:u-1]\Big) 
\\&+ I\Big(X_K; |h_{\rm d}| X_K + Z_K \Big| X_\ell, \ h_{\rm i} X_K + Z_\ell, \ \ell\in[1:K-1]\Big)  
  %
\\& \leq h\Big( |h_{\rm d}| X_1  + h_{\rm i} \sum_{i=2}^{K}X_i + Z_1\Big)-h(Z_1)
\\&+ \sum_{u=2}^{K-1}h\Big([|h_{\rm d}|-h_{\rm i}] X_u + Z_u-Z_{u-1}) -h(Z_u) 
\\&+ h\Big(|h_{\rm d}| X_K + Z_K \Big|
  h_{\rm i} X_K +  \frac{1}{K-1}\sum_{\ell=1}^{K-1}Z_\ell\Big) -h(Z_K).
\end{align*}
Finally, by the ``Gaussian maximizes entropy'' principle, we obtain
\begin{subequations}
\begin{align}
\sum_{k=1}^{K} R_k
  &\leq   \log\left(1+\Big(|h_{\rm d}|+(K-1)|h_{\rm i}|\Big)^2   \right)\label{eq:up awgn sym first}
\\&+(K-2)\log(2) + (K-2) \log\left(1+\frac{\big||h_{\rm d}|-h_{\rm i}\big|^2}{2}\right)\label{eq:up awgn sym middle}
\\&+      \log\left(1+\frac{|h_{\rm d}|^2}{1+(K-1)|h_{\rm i}|^2} \right)\label{eq:up awgn sym last}.
\end{align}
\label{eq:up awgn sym}
\end{subequations}
For $h_{\rm i}=|h_{\rm d}|$ all received signals are statistically equivalent, therefore the $K$-CIFC-CMS is equivalent to a $K$-user Multiple Access Channel, whose sum-capacity is
\begin{align*}
\sum_{k=1}^{K} R_k 
  &\leq I(X_1,\ldots,X_K; |h_{\rm d}| \sum_{i=1}^{K}X_i + Z_1)
\\&\leq \log(1+K^2 |h_{\rm d}|^2).
\end{align*}

In the limit for high SNR and with the channel parameterization as in~\eqref{eq:ch parm}, the above upper bound can be further bounded
\begin{align*}
\sum_{k=1}^{K} R_k 
  &\leq \log(K^2) +(K-1)\log(2)+ (K-1)\log\Big(1+\max\{|h_{\rm d}|^2,|h_{\rm i}|^2\}\Big)
\\&+ \log\left(1+\frac{|h_{\rm d}|^2}{1+(K-1)|h_{\rm i}|^2}\right).
\end{align*}
to obtain the following gDoF upper bound 
\begin{align*}
d(\alpha) \leq (K-1)\max\{1,\alpha\}+[1-\alpha]^+ = K\max\{1,\alpha\}-\alpha.
\end{align*}
This gDoF remain valid for $\alpha=1$ as long as $h_{\rm i}=|h_{\rm d}|\exp(j \theta)$
for $\exp(j \theta)\not=1$; when $\exp(j \theta)=1$ the $K$-user MAC sum-capacity gives $d(\alpha=1) = 1$. 
This proves the converse part of Th.~\ref{thm:K-AWGN-dgdof}.

\subsection{Achievable Rate Region for K-CIFC with CMS}
\label{sec:K-AWGN-sym-in}

In this section we describe a scheme which we shall in the Section~\ref{sec:K-AWGN-sym-gap} to show that the symmetric upper bound derived in Section~\ref{sec:K-AWGN-sym-out} is achievable to within a constant gap.

Inspired by the capacity achieving strategy for the Gaussian MIMO-BC, we introduce a scheme that uses Dirty Paper Coding (DPC) with encoding order $1\rightarrow 2\rightarrow 3 \rightarrow \cdots K$.
We denote by $\mathbf{\Sigma}_\ell$ the covariance matrix corresponding to the message intended for decoder $\ell$,  $\ell\in[1:K]$, as transmitted across the $K$ antennas/transmitters. The overall input covariance matrix is
\begin{subequations}
\begin{align}
{\rm Cov}[X_1,\ldots,X_K]
= \sum_{\ell=1}^{K} \mathbf{\Sigma}_\ell :  
\left[\sum_{\ell=1}^{K} \mathbf{\Sigma}_\ell\right]_{k,k} \leq 1, \ k\in[1:K],
\end{align}
where the constraints on the diagonal elements correspond to the input power constraints. Moreover, since message~$\ell$ can only be broadcasted by transmitters with index larger than $\ell$, we further impose
\begin{align}
\big[\mathbf{\Sigma}_\ell\big]_{k,k} = 0 \ \text{for all $1\leq k < \ell \leq K$}.
\end{align}
\label{eq:mimobclike cov constraints}
\end{subequations}
The achievable rate region is then the set of non-negative rates $(R_1,\ldots,R_K)$ that satisfy
\begin{align}
R_\ell \leq \log\left(1+\frac{\mathbf{h}_\ell^\dagger \mathbf{\Sigma}_\ell \mathbf{h}_\ell}{\mathbf{h}_\ell^\dagger \left(\sum_{k=\ell+1}^{K}\mathbf{\Sigma}_k\right) \mathbf{h}_\ell}\right), \
\mathbf{h}_\ell^\dagger := [h_{\ell,1} h_{\ell,2} \ldots h_{\ell,K}], \ \ell\in[1:K],
\label{eq:mimobclike rates}
\end{align}
for all possible ${\rm Cov}[X_1,\ldots,X_K]$ complying with~\eqref{eq:mimobclike cov constraints}, with the convention that $\sum_{k=K+1}^{K}\mathbf{\Sigma}_k = 0$. 

In particular we consider the transmit signals
\begin{align*}
  &X_1 =                                                            \alpha_1 U_1,
\\&X_j = \gamma_{j} U_{j} + \beta_j                 U_j^{\rm(ZF)} + \alpha_j U_1, \ j\in[2:K-1],
\\&X_K = \gamma_{K} U_{K} - \beta_K \sum_{j=2}^{K-1}U_j^{\rm(ZF)} + \alpha_K U_1,
\end{align*}
where $U_\ell, U_\ell^{\rm(ZF)} \ \text{are i.i.d.} \ \mathcal{N}(0,1),  \ell\in[1:K],$
and the coefficients $\{\alpha_1, \alpha_j,\beta_j,\gamma_j\}_{j\in[2:K]}$ are such that
\begin{align*}
  & |\alpha_1|^2\leq 1,
\\& |\gamma_{j}|^2+|\beta_j|^2+|\alpha_j|^2\leq 1, \ j\in[2:K-1],
\\& |\gamma_{K}|^2+|\beta_K|^2(K-2)+|\alpha_K|^2\leq 1,
\end{align*}
in order to satisfy the power constraints.
Notice the negative sign for $\beta_K$, which we shall use to implement zero-forcing of the aggregate interference $\sum_{j=2}^{K-1}U_j^{\rm(ZF)}$. Moreover, all transmitters cooperate in beam forming $U_1$ to receiver~1. These two facts can be easily seen by observing that for  $\beta_1=\ldots=\beta_K:=\beta$
\begin{align*}
 \left. \sum_{\ell=1}^{K} X_\ell \right|_{\beta_1=\ldots=\beta_K}
= \sum_{\ell=1}^{K} \gamma_\ell U_\ell , \ \  {\gamma_1:=\sum_{\ell=1}^{K}\alpha_\ell}.
\end{align*}

With these choices the message covariance matrices are
\begin{align*}
\mathbf{\Sigma}_1    &= \mathbf{a} \mathbf{a}^\dagger, \ \mathbf{a} := [\alpha_1,\ldots,\alpha_K]^T,
\\
\mathbf{\Sigma}_j &= |\gamma_j|^2 \ \mathbf{e}_j \mathbf{e}_j^\dagger  
+|\beta|^2 \ (\mathbf{e}_j-\mathbf{e}_K)(\mathbf{e}_j-\mathbf{e}_K)^\dagger,
 \ j\in[2:K],
\end{align*}
where $\mathbf{e}_j$ indicates a length-$K$ vector of all zeros except for a one in position $j$, $j\in[1:K]$,
$\dagger$ indicates the Hermitian transpose, and where  $\beta=\beta_1=\ldots=\beta_K$.

We next express the channel vectors $\mathbf{h}_\ell$ for the symmetric Gaussian channel as
\[
\mathbf{h}_\ell 
=  (|h_{\rm d}|-h_{\rm i}) \ \mathbf{e}_\ell
+ h_{\rm i} \ \left(\sum_{k=1}^{K}\mathbf{e}_k\right), \ \ell\in[1:K].
\]
By noticing that
\[
\mathbf{h}_\ell \mathbf{e}_j^\dagger
= \delta[\ell-j] (|h_{\rm d}|-h_{\rm i})+ h_{\rm i},
 \ \ell\in[1:K],
\]
where $\delta[k]$ is the Kronecker's delta function, the following achievable rates are achievable
\begin{subequations}
\begin{align}
   R_1 &= \log\left( 1+\frac{\Big| |h_{\rm d}|+|h_{\rm i}|\sum_{j=2}^{K}\alpha_j \Big|^2}{1+|h_{\rm i}|^2\sum_{k=2}^{K}|\gamma_k|^2}\right), \label{eq:ach awgn sym first}
\\ R_j &= \log\left(1+\frac{ \Big| |h_{\rm d}|-h_{\rm i} \Big|^2 \ |\beta|^2+|h_{\rm d}|^2 \ |\gamma_j|^2}{1+|h_{\rm i}|^2\sum_{k=j+1}^{K}|\gamma_k|^2}\right), j \in [2:K-2], \label{eq:ach awgn sym middle}
\\ R_K &= \log\left(1+|h_{\rm d}|^2 \ |\gamma_K|^2 \right), \label{eq:ach awgn sym last}
\end{align}
\label{eq:ach awgn sym}
\end{subequations}
where we chose $\alpha_1=\exp(j \angle{h_{\rm i}})$ (notice the phase of $\alpha_1$ which allows coherent combining at receiver~1 of the different signals carrying $U_1$, i.e., all users beamform to the primary receiver).

\subsection{Additive Constant Gap Results for the symmetric Gaussian Noise Channel}
\label{sec:K-AWGN-sym-gap}

We now choose the parameters in~\eqref{eq:ach awgn sym} so as to match the upper bound in~\eqref{eq:up awgn sym}. 

A tempting interpretation for the bound in~\eqref{eq:up awgn sym last} is to say that the most cognitive user should treat all the other signals as noise, because of the term $(K-1)|h_{\rm i}|^2$ at the denominator of the equivalent SNR for receiver~$K$. However we recall that user~$K$ is the most cognitive user and can therefore `pre-code' the whole interference seen at its receiver by using DPC; by doing so, receiver~$K$ would not have anything to treat as noise besides the Gaussian noise itself. We therefore interpret the term $\frac{1}{1+(K-1)|h_{\rm i}|^2} \leq 1$ as the fraction of power transmitter~$K$ dedicates to the transmission of its own signal. This amounts to setting 
\begin{align*}
|\gamma_K|^2 = \frac{1}{1+(K-1)|h_{\rm i}|^2}
\end{align*}
in~\eqref{eq:ach awgn sym last}. This choice guarantees that the achievable rate for user~$K$ exactly matches the term in~\eqref{eq:up awgn sym last} in the upper bound.

Next we would like to match the upper bound term in~\eqref{eq:up awgn sym middle} to the achievable rates in~\eqref{eq:ach awgn sym middle} by setting
\[
\gamma_j=0, \ j\in[2,K-1],
\quad
\frac{1}{2} = \frac{|\beta|^2}{1+|h_{\rm i}|^2|\gamma_K|^2}.
\]
However, from the power constraint for user~$K$, we must satisfy 
\[
|\beta|^2 \leq \frac{1-|\gamma_K|^2}{K-2},
\]
which imposes the following condition
\[
\frac{K-4}{K-2}+\left(|h_{\rm i}|^2+\frac{2}{K-2}\right) |\gamma_K|^2\leq 0.
\]
The above condition cannot be satisfied for $K\geq 4$;
for $K=3$ it requires that
\[
|\gamma_3|^2 = \frac{1}{1+2|h_{\rm i}|^2} \leq \frac{1}{|h_{\rm i}|^2+2}
\]
which can be satisfied by $|h_{\rm i}|^2 \geq 1$.
Therefore, in the following we shall assume $|h_{\rm i}|^2 \geq 1$
and set $\gamma_j=0, \ j\in[2,K-1]$ and
\[
|\beta|^2
= |\beta_2|^2=\ldots=|\beta_K|^2 
= \left\{\begin{array}{ll}
\frac{1-|\gamma_K|^2}{K-2}=\frac{1}{K-2}\left( 1- \frac{1}{1+(K-1)|h_{\rm i}|^2} \right) & \ K\geq 4 \\
\frac{1+|h_{\rm i}|^2|\gamma_3|^2}{2} = \frac{1+3|h_{\rm i}|^2}{2(1+2|h_{\rm i}|^2)} & \ K= 3 \\
\end{array}\right.,
\]
which implies
\[
|\alpha_K|^2  
= \left\{\begin{array}{ll}
0 & \ K\geq 4 \\
1-|\beta|^2-|\gamma_K|^2=\frac{-1+|h_{\rm i}|^2}{2(1+2|h_{\rm i}|^2)} & \ K= 3 \\
\end{array}\right..
\]
Finally, for $j\in[2:K-1]$
\[
|\alpha_j|^2 
= 1- |\beta_j|^2 
= \left\{\begin{array}{ll}
 \frac{K-3}{K-2}+\frac{1}{K-2} \ \frac{1}{1+(K-1)|h_{\rm i}|^2} & \ K\geq 4 \\
 \frac{1+|h_{\rm i}|^2}{2(1+2|h_{\rm i}|^2)} & \ K= 3 \\
\end{array}\right..
\]

The rates then become: for $K\geq 4$
\begin{align*}
   R_K &= \log \left(1+ \frac{|h_{\rm d}|^2}{1+(K-1)|h_{\rm i}|^2}\right)
\\ R_j &= \log \left(1+  \big||h_{\rm d}|-h_{\rm i} \big|^2 
\frac{ \frac{1}{K-2}  \ \frac{(K-1)|h_{\rm i}|^2}{1+(K-1)|h_{\rm i}|^2} }{1+\frac{|h_{\rm i}|^2}{1+(K-1)|h_{\rm i}|^2}}\right), 
\\&\geq  \log \left(1+  \frac{\big||h_{\rm d}|-h_{\rm i} \big|^2}{K-2}  \frac{K-1}{K+1}\right),\ j\in[2:K-1], \text{since $|h_{\rm i}|^2\geq 1$},
\\ R_1 &= 
\log \left(1+ \frac{\left| |h_{\rm d}|+|h_{\rm i}|  \sqrt{(K-3)(K-2)+\frac{K-2}{1+(K-1)|h_{\rm i}|^2}} \right|^2}{1+\frac{|h_{\rm i}|^2}{1+(K-1)|h_{\rm i}|^2}}\right) 
\\& \geq \log \left(1+ \frac{\big||h_{\rm d}|+|h_{\rm i}|\sqrt{(K-3)(K-2)}\big|^2}{2} \right), \text{since $|h_{\rm i}|^2\geq 1$},
\end{align*}
and for $K=3$
\begin{align*}
   R_3 &= \log \left(1+ \frac{|h_{\rm d}|^2}{1+2|h_{\rm i}|^2}\right)
\\ R_2 &= \log \left(1+ \big||h_{\rm d}|-h_{\rm i} \big|^2  \frac{1}{2} \right)
\\ R_1 &= \log \left(1+ \frac{\left| |h_{\rm d}|+|h_{\rm i}| \ \left(
 \sqrt{\frac{ 1+|h_{\rm i}|^2}{2(1+2|h_{\rm i}|^2)}}
+\sqrt{\frac{-1+|h_{\rm i}|^2}{2(1+2|h_{\rm i}|^2)}}
\right) \right|^2}
{1+\frac{|h_{\rm i}|^2}{1+2|h_{\rm i}|^2}} \right) 
\\&\geq \log \left(1+\frac{\left| |h_{\rm d}|+|h_{\rm i}|\frac{1}{2}\right|^2}{2} \right), \ \text{since $|h_{\rm i}|^2\geq 1$}.
\end{align*}

By taking the difference between the upper bound in~\eqref{eq:up awgn sym} and the derived achievable-rates we find that the gap is upper bounded by: for $K\geq 4$
\begin{align*}
\mathsf{GAP}
& \leq (K-2)\log(2) + (K-2)\left(
\log\left(1+\frac{\big||h_{\rm d}|-h_{\rm i}\big|^2}{2}\right)
- \log \left(1+  \frac{\big||h_{\rm d}|-h_{\rm i} \big|^2}{K-2}  \frac{K-1}{K+1}\right)
\right)
\\&+ \log\left(1+\Big(|h_{\rm d}|+(K-1)|h_{\rm i}|\Big)^2   \right)
   - \log \left(1+ \frac{\big||h_{\rm d}|+|h_{\rm i}|\sqrt{(K-3)(K-2)}\big|^2}{2} \right)
\\&\leq  (K-2) \log(2) +  (K-2)\log\left(\frac{(K+1)(K-2)}{2(K-1)}\right)
+ \log\left(\frac{2(K-3)(K-2)}{(K-1)^2}\right)
\\& \leq (K-2)\log\left(K-2\right) +\log(2\exp(2)),
\end{align*}
(where we used $K\log_{\rm e}(1+1/K)\leq 1$)
and for $K\geq 3$
\begin{align*}
\mathsf{GAP}
&\leq \log(2)+\log\left(1+\Big(|h_{\rm d}|+2|h_{\rm i}|\Big)^2   \right)
 -    \log \left(1+\frac{\left| |h_{\rm d}|+|h_{\rm i}|\frac{1}{2}\right|^2}{2} \right)
\\&\leq 6\log(2).
\end{align*}

For $|h_{\rm i}|^2<1$, we can set $\beta_j=\alpha_j=0, \gamma_j=1$ for $j\in[2:K]$ to obtain
\[
\sum_{\ell=1}^{K} R_\ell = 
\sum_{\ell=1}^{K} \log\left(1+\frac{|h_{\rm d}|^2}{1+(K-\ell)|h_{\rm i}|^2}\right).
\]
The gap to the upper bound is at most
\begin{align*}
\mathsf{GAP}
\leq  (K-2)\log(2)+2\log(K-1)+\sum_{\ell=2}^{K-1}\log\left(\frac{K-\ell}{2}\right).
\end{align*}
which is smaller than the gap previously obtained for $|h_{\rm i}|^2\geq 1$. 

This proves Th.~\ref{thm:K-AWGN-sym-add} and implies the direct part of Th.~\ref{thm:K-AWGN-dgdof}.

\subsection{Multiplicative Constant Gap Results for the symmetric Gaussian Noise Channel}
\label{sec:K-AWGN-sym-muliplicative}

In order to provide a complete characterization of the sum-capacity of the symmetric Gaussian channel we next consider approximate the sum-capacity to within a multiplicative gap, more relevant at low SNR than additive gaps. 
%
%
%
To this end, note that the rate of user $j$ is upper bounded by $C_j : = \log(1+ (|h_{\rm d}| + (K-j)|h_{\rm i}|)^2), j\in [1:K]$
 which in turn is upper bounded by $K \times C_1$. Consider an achievability scheme in which all users beamform to user 1: this achieves the 
 sum-rate $R_1+\cdots R_K = C_1$. This is to within a factor $K$ of the upper bound, proving Th.~\ref{thm:K-AWGN-sym-mul}. 

%
%
%
%
%

\subsection{Numerical optimization of inner and outer bounds for the symmetric 3-user case }
\label{sec:K-AWGN-sym-num}

\begin{figure}
\centering
\includegraphics[width=10cm]{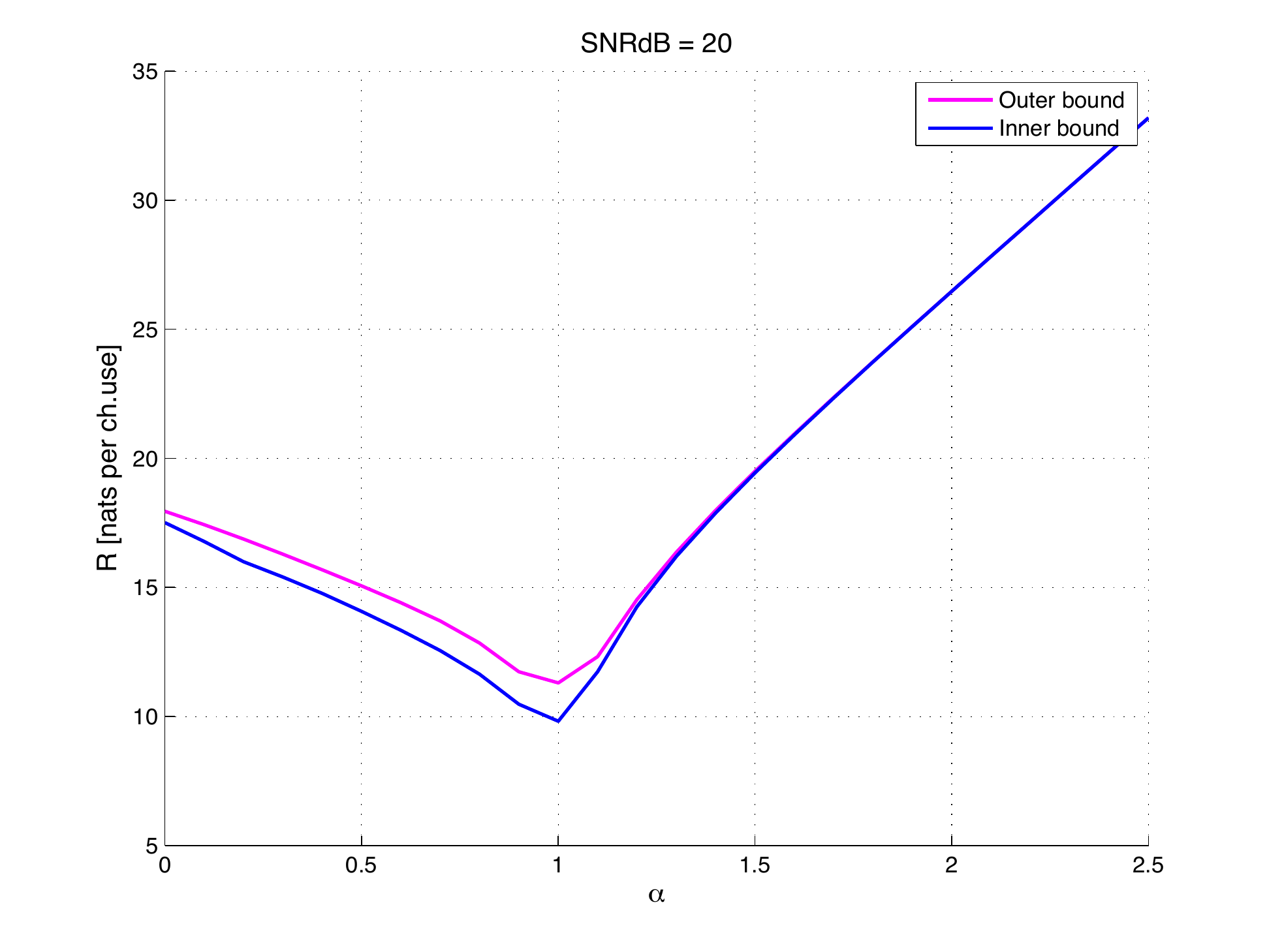}
\caption{Comparison of the numerically optimized inner and outer bounds for $K=3$ users at SNR$=20$dB as a function of $\alpha = \frac{\log(|h_{\rm d}|)}{\log(|h_{\rm i}|)}$; notice a smaller gap than the worst case predicted 6 bits per channel use per user.}
\label{fig:numcomp}
\end{figure}

\begin{figure}
\centering
\includegraphics[width=10cm]{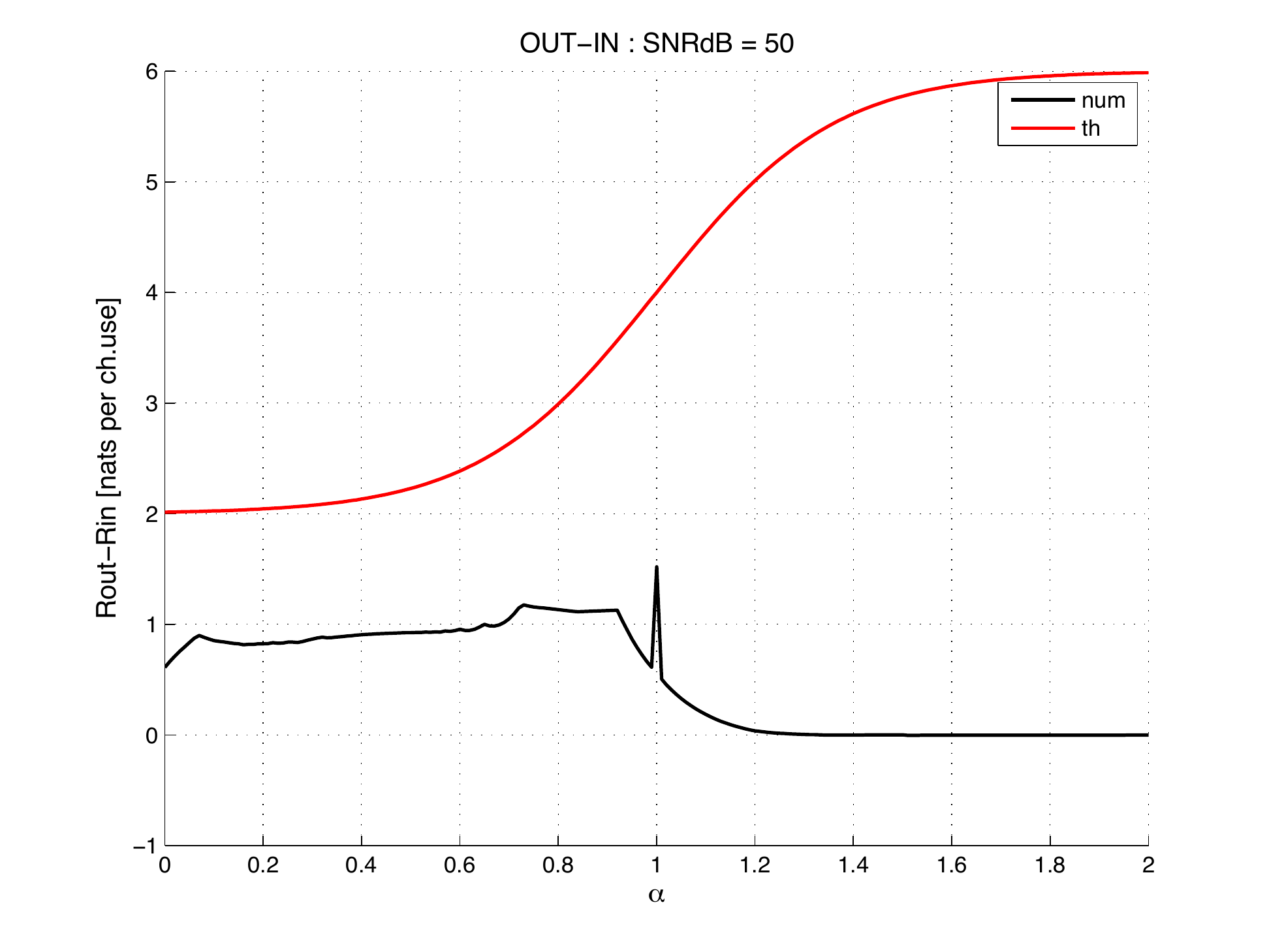}
\caption{Analytical and numerical additive gaps for $K=3$ users at SNR$=50$dB.}
\label{fig:numcomp2}
\end{figure}

Fig.~\ref{fig:numcomp} shows the proposed upper and lower bound for the symmetric channel with $K=3$ users and SNR$=20$dB. In this case the upper and lower bounds where optimized numerically so as to obtain a larger achievable rate and a tighter outer bound than those used for the analytical evaluation of the gap. We notice that the gap between the bounds is much less than the theoretical gap of 6 bits. In particular, for strong interference the bounds are extremely close to one another, showing again that the theoretical gap of 6 bits is a worst case scenario. { Fig.~\ref{fig:numcomp2} shows the additive gap for $K=3$ users at SNR$=50$dB; notice the gap between the analytical upper and lower bounds (curve labeled `th') converging to 6~bits for large $\alpha$ while the gap between the numerically optimized upper and lower bounds (curve labeled `num') going to zero in the same regime; the larger gap is at $\alpha=1$ where the channel matrix becomes rank deficient; overall the gap is at most of the order of 1~bit, which is about 5 bits smaller than the analytical gap.} 
%
%
%
%
%
%

\section{Conclusion}
\label{sec:conclusion}

In this paper we studied the $K$-user cognitive interference channel with cumulative message sharing.
A computable, general outer bound valid for any number of users and any memoryless channel is obtained.
For the linear deterministic approximation of the Gaussian channel at high SNR we obtained the sum-capacity for all channel gains in the case of three users, and the symmetric sum-capacity for any $K$. 
For the Gaussian channel, we provided a unified achievability scheme which achieves to within a constant additive and multiplicative gap the sum-capacity outer bound. 
In the linear deterministic channel, the sum-capacity was achieved by a scheme which only required cognition at one single user. This begs the question of whether, for the Gaussian channel, one may achieve to within a constant gap of capacity by only having one fully cognitive user; our current achievability scheme does require cognition at intermediate transmitters for dirty paper coding. 
Furthermore, comparisons with different $K$ user cognitive models 
are of interest and subject of current investigation.


\section*{Acknowledgment}
The work of the authors was partially funded by NSF under awards 0643954 and 1017436.
The contents of this article are solely the responsibility of the authors and do not necessarily
represent the official views of the NSF.

\appendices

\section{Proof of Th.~\ref{thm:outer K=3}}
\label{app:proof of thm:outer K=3}

By Fano's inequality $H(W_i|Y_i^N) \leq N \epsilon_N$ with $\epsilon_N\to 0$ as $N\to\infty$
for all $i \in [1:3]$.

The bounds in equation~\eqref{eq:K=3 1} through~\eqref{eq:K=3 3} are a simple application of the cut-set bound.

The bound in~\eqref{eq:K=3 2,3} is obtained as follows:
\begin{align*}
  &N(R_2+ R_3- 2 \epsilon_N) 
 \stackrel{\rm(a)}{\leq} I(Y_2^N;W_2)    +I(Y_3^N;W_3)
\\& \stackrel{\rm(b)}{\leq} I(Y_2^N,W_1;W_2)+I(Y_3^N,Y_2^N,W_1,W_2;W_3)
\\& \stackrel{\rm(c)}{=   } I(Y_2^N;W_2|W_1)+I(Y_3^N,Y_2^N;W_3|W_1,W_2)
\\& \stackrel{\rm(d)}{=   } I(Y_2^N;W_2|W_1)+I(Y_2^N;W_3|W_1,W_2)
\\&\qquad +I(Y_3^N;W_3|W_1,W_2,Y_2^N)
\\& \stackrel{\rm(e)}{=   } I(Y_2^N;W_2,W_3|W_1) + I(Y_3^N;W_3|W_1,W_2,Y_2^N)
\\& \stackrel{\rm(f)}{=   } I(Y_2^N;W_2,W_3|W_1,X_1^N)
\\&\qquad  + I(Y_3^N;W_3|W_1,W_2,Y_2^N,X_1^N,X_2^N)
\\& \stackrel{\rm(g)}{\leq} \sum_{t=1}^{N} H(Y_{2,t}|X_{1,t})-H(Y_{2,t}|X_{1,t},X_{2,t},X_{3,t})
\\&\quad + H(Y_{3,t}|X_{1,t},X_{2,t})-H(Y_{3,t}|X_{1,t},X_{2,t},X_{3,t})
\\& \stackrel{\rm(h)}{=   }\sum_{t=1}^{N} I(Y_{2,t};X_{2,t},X_{3,t}|X_{1,t})+I(Y_{3,t};X_{3,t}|X_{1,t},X_{2,t}),
\end{align*}
where
(a) follows from Fano's inequality, 
(b) the non-negativity of mutual information, 
(c) from the independence of the messages,
(d) and (e) from chain rule (note how we gave side information so that we could recombine different entropy terms), 
(f) because the inputs are deterministic functions of the messages,
(g) follows since conditioning does not reduce entropy, and
(h) definition of mutual information.

With similar steps (give enough messages so that we can reconstructs the inputs as also give
outputs so that we can recombine terms by using the chain rule of mutual information)
we obtain the bound in~\eqref{eq:K=3 1,2,3}. The main steps are:
\begin{align*}
  &N(R_1+R_2+R_3- 3 \epsilon_N) \\&\leq I(Y_1^N; W_1) + I(Y_2^N; W_2) +I(Y_3^N; W_3)
\\&\leq I(Y_1^N; W_1) + I(Y_2^N, Y_1^N,W_1;W_2)+I(Y_3^N,Y_1^N,W_1,Y_2^N,W_2;W_3)
\\&\leq I(Y_1^N; W_1,W_2,W_3)  + I(Y_2^N;W_2,W_3|Y_1^N,W_1)+I(Y_3^N;W_3|Y_1^N,W_1,Y_2^N,W_2)
\\&\leq \sum_{t=1}^{N} I(Y_{1,t}; X_{1,t},X_{2,t},X_{3,t})   + I(Y_{2,t}; X_{2,t},X_{3,t}| X_{1,t},Y_{1,t})    + I(Y_{3,t}; X_{3,t}        | X_{1,t},X_{2,t}, Y_{1,t},Y_{2,t}).
\end{align*}

\section{Proof of Th.~\ref{thm:outer K}}
\label{app:proof of thm:outer K}

By Fano's inequality $H(W_i|Y_i^N) \leq N \epsilon_N$ with $\epsilon_N\to 0$ as $N\to\infty$
for all $i \in [1:K]$.

For~\eqref{eq:outer bound general K single} we have
\begin{align*}
  &N(R_i-\epsilon_N) \\&\leq I(Y_i^N; W_i)
\\&\leq I(Y_i^N; W_i|W_1,\ldots, W_{i-1})
\\&=    \sum_{t=1}^{N} h(Y_{i,t}|W_1,\ldots, W_{i-1}, Y_i^{t-1}) - h(Y_{i,t}|W_1,\ldots, W_{i}, Y_i^{t-1})
\\&\leq \sum_{t=1}^{N} h(Y_{i,t}|W_1,\ldots, W_{i-1}) - h(Y_{i,t}|W_1,\ldots, W_K, Y_i^{t-1})
\\&\leq \sum_{t=1}^{N} h(Y_{i,t}|X_{1,t},\ldots, X_{i-1,t}) - h(Y_{i,t}|X_{1,t},\ldots, X_{K,t})
\\&=    \sum_{t=1}^{N} I(Y_{i,t}; X_{i,t},\ldots, X_{K,t}|X_{1,t},\ldots, X_{i-1,t}).
\end{align*}

For~\eqref{eq:outer bound general K progressive} we have
\begin{align*}
  &N\sum_{j=i}^{K}(R_j-\epsilon_N)
\\&\leq \sum_{j=i}^{K}I(Y_j^N; W_j)
\\&\leq \sum_{j=i}^{K}I(Y_j^N,  
   \ \underbrace{W_1,\ldots,W_{i-1}}_{=\emptyset \ \text{for $i=1$}}, 
   \ \underbrace{Y_i^N, W_i, \ldots Y_{j-1}^N, W_{j-1}}_{=\emptyset \ \text{for $j=i$}};
   W_j)
\\&=    \sum_{j=i}^{K}I(Y_i^N,\ldots Y_j^N; W_j | W_1,\ldots,W_{j-1})
\\&=    \sum_{j=i}^{K}\sum_{k=i}^{j}I(Y_k^N; W_j | W_1,\ldots,W_{j-1}, \ Y_i^N,\ldots, Y_{k-1}^N)
\\&=    \sum_{k=i}^{K}\sum_{j=k}^{K}I(Y_k^N; W_j | W_1,\ldots,W_{j-1}, \ Y_i^N,\ldots, Y_{k-1}^N)
\\&=    \sum_{k=i}^{K}I(Y_k^N; W_k,\ldots,W_K | \underbrace{W_1,\ldots,W_{i-1}}_{=\emptyset \ \text{for $i=1$}},\ \underbrace{W_i,Y_i^N,\ldots, W_{k-1},Y_{k-1}^N}_{=\emptyset \ \text{for $k=i$}})
\\&\leq \sum_{k=i}^{K}\sum_{t=1}^{N}  I(Y_{k,t};X_{k,t},\ldots,X_{K,t}  | X_{1,t},\ldots,X_{k-1,t}, Y_{i,t},\ldots,Y_{k-1,t}).
\end{align*}

\bibliography{refs}
\bibliographystyle{IEEEtran}

\newpage


\end{document}